# "Is a picture of a bird a bird":
# Policy recommendations for dealing with ambiguity in machine vision models

Alicia Parrish, Sarah Laszlo, and Lora Aroyo
*Google Research*

**Abstract**

Many questions that we ask about the world do not have a single clear answer, yet typical human annotation set-ups in machine learning assume there must be a single ground truth label for all examples in every task. The divergence between reality and practice is stark, especially in cases with inherent ambiguity and where the range of different subjective judgments is wide. Here, we examine the implications of subjective human judgements in the behavioral task of labeling images used to train machine vision models. We identify three primary sources of ambiguity arising from (i) depictions of labels in the images, (ii) raters' backgrounds, and (iii) the task definition. On the basis of the empirical results, we suggest best practices for handling label ambiguity in machine learning datasets.

**Introduction**

Much of the policy guidance in the field of artificial intelligence is at a broad, application and model agnostic, conceptual level. For example, the NIST standards for AI risk management (Tabassi, 2023) are scoped to "an engineered or machine-based system that can, for a given set of objectives, generate outputs such as predictions, recommendations, or decisions influencing real or virtual environments." Similarly, the UNESCO AI and Education policy guidance (Miao, 2021) defines AI as "computer systems that have been designed to interact with the world through capabilities that we usually think of as human.", quoting (Luckin, 2017). We see this domain generality also in the FDA's regulatory framework for AI and ML based software (FDA, 2019), which is scoped to AI as "the science and engineering of making intelligent machines, especially intelligent computer programs," citing McCarthy (2007).

Policy guidance for the use of computer vision models more specifically is more limited, and seems to cluster around a small number of computer vision applications that are likely to serve or impact the public good. These are, especially, the use of computer vision in medicine (e.g., Harned, Lungren, & Rajpurkar, 2019), city planning (Kang et al., 2020), and criminal justice (discussed in Margetts, 2022).

Even within the more limited literature making policy recommendations for the use of machine vision specifically, there is little to no work which provides recommendations for policies pertaining to *large label space* models, which are the focus of the current work. Large label space models are a specific type of machine vision model which, when presented with an image as input, produce as output the probabilities that each of a large number of entities are present in the input image (review in Bogatinovsky, 2022). This is in contrast to *binary classification* models, which might only say whether a single entity is or is not present in an image, or *segmentation models*, which might identify the pixels corresponding to an entity in an image. As an illustrative example, a large label space model might say, for a picture of an animal, how likely it is that the animal is a member of each species of *Aves* (birds). A binary classification model might instead say whether the image has a bird in it. And, a segmentation model might identify the pixels in the image that make up the animal.

Here, we focus on large label space models.  Most (but not all, see, for example, Ji, Henriques, & Vedaldi, 2019) image classification models require labeled training data in order to learn to perform classification accurately.  This typically consists of a *training set* of images which are labeled-- usually by human annotators-- with respect to their contents.  In the bird example, in order to learn to be able to classify a bird by its species, a large label space model would need to see many (usually at least tens of thousands) of images of birds labeled for their species.  Human annotators, who know which birds are members of which species, might be employed to label each image, and provide the "ground truth" needed to train the model.  Indeed, in one example of computer vision applied towards public health, entomologists were employed to label images of mosquitos taken by citizen scientists, with the goal of enabling better tracking of disease outbreaks (Munoz et al., 2020).

What happens, however, when human raters disagree?  What if the bird experts do not agree on which species an image belongs to?  We can imagine many reasons that this might happen-- the image might not be good enough to make an assessment, there may be multiple species that are very similar in appearance, names of birds might be different in one geographical location than another, or the raters might have different understandings of the purpose of the task-- for example, some might label images according to genus and another to species.  There are, then, many possible types of label ambiguity, and this is a challenge because label ambiguity is typically deleterious to machine vision models (see, for example, Karimi et al., 2020, for an analysis of the impact of label noise on medical image analysis models).  The challenge arries because the model can only learn as well as its data, and ambiguous data will lead to ambiguous model predictions.  Thus, understanding label ambiguity is part and parcel to improving machine vision models.

Towards this goal--to better understand the impact of label ambiguity on large label space model performance--we conceived and implemented a data challenge: CATS4ML (Crowdsourcing Adverse Test Sets for Machine Learning). In this open challenge, available online, participants from the global research community competed by attempting to improve the data available for large label space model training[1]. The challenge required participants to identify images that state of the art image-classification models might be incorrectly classifying.  The goal was to understand systematic failures of state of the art models, with an eye towards augmenting the data used to train these models in the future to better cover the failure cases.  All image-label pairs collected during the challenge were tested against multiple state of the art classification models and surfaced many pairs with clear human-machine disagreements and even pairs where multiple human annotators couldn't reach a clear agreement with each other.

Here, we describe the results of the challenge.  We analyze in detail the ambiguities present in the data, and organize them into a theoretical framework.  We then make recommendations for human annotation and data collection policies that best address the types of ambiguities we observed.

**CATS4ML Challenge**

The CATS4ML challenge ran online from January-April 2021, under the Crowd Camp umbrella of the Human Computation Conference.  The dataset that the Challenge used as source material was the Open Image Dataset (OID), V4[2].  The OID V4 (Krasin et. al 2017) consists of ~9M images annotated with 20k possible image-level labels (e.g., "this image includes a bird"), object bounding boxes (e.g., "this is a box around the portion of the image that includes the bird") and segmentation masks (e.g., "these are the pixels that make up the bird").  Importantly, the labels, bounding boxes, and segmentation masks in the OID are provided by a machine, not by a person. CATS4ML was designed under the premise that, likely, the machine labeler makes mistakes, that these mistakes are likely systematic, and that studying systematic machine failures can improve the machine labelers in the future.

---

[1] https://ai.googleblog.com/2021/02/uncovering-unknown-unknowns-in-machine.html
[2] https://storage.googleapis.com/openimages/web/factsfigures_v4.html

Participants were instructed to examine OID images that were machine-labeled as including the entities Bird, Canoe, Lipstick, Chopsticks, Muffin, Pizza, Croissant, Child, Smile, Selfie, American football, Athlete, Physician, Nurse, Teacher, Chef, Firefighter, Coach, Construction worker, Bus driver, Funeral, Thanksgiving, or Graduation, and submit image-label pairs where they thought that the image classification machine algorithm could be wrong. It was necessary to limit the label set under investigation in order to make the scope of the competition tractable-- it would not have been realistic to expect human participants to examine all 20k labels present in the OID. These particular 23 labels were selected to represent a neutral (non-controversial, non-sensitive) set of topics across different types: objects (8), events (3), roles and professions (9), and abstract concepts (3). Another criteria for selection was to have a good representation of different levels of ambiguity of the label, e.g. "child" is a broad concept and could be interpreted in many different ways; "athlete" could mean different things for different cultures; "physician" and "nurse" could be very ambiguous in their visual representation.

Within the set of images designated by the OID as including these 23 labels, participants were instructed to identify images that they thought might be challenging for the machine-labeller-- that is, images where the machine-labeller was likely, in their human estimation, to make a mistake.

10 individuals submitted image-label pairs to the challenge. Two challenge participants self-identified as coming from industry, and 8 from academia. The participants were global, self-identifying as being from Japan (1), Australia (2), the United States (4), India (2), and The Netherlands (1). These 10 individuals submitted more than 14,000 image-label pairs. Of these, 13,683 pairs were "valid", where "valid" meant that the pairs were in fact drawn from the 23 challenge labels. Of the valid pairs, only 10,668 pairs were unique.

The 10,668 unique image-label pairs were further validated by engaging two globally-diverse crowds of raters in different locales and two in-house experts in three human annotation tasks. This process is described in detail in the Methods. The image-label pairs were also submitted to six machine vision models, in order to examine how human judgements aligned with state of the art model judgements. The original CATS4ML participants, who submitted the questioned image-label pairs in the first place, were, at the end of the experimental procedure, assigned a position on a public leaderboard based on how many errors in the OID labels they had identified, as indicated by the human and state-of-the-art model consensus.

**Methods**
**Participants**
The data submitted by challenge participants was thrice validated-- twice by research participants and once by internal confederates. External participants for the validation were recruited from professional rater pools. Raters in these pools are trained and experienced in various trust and safety tasks. We engaged two non-overlapping globally-diverse rater pools. We did not collect demographic information aside from locale for these participants.

The two non-overlapping groups of participants labeled images in 2 separate tasks. These are described in detail below, but, in brief, Task 1 required participants to verify whether or not a particular label was present in an image, where the labels and images were the image/label pairs submitted by the CATS4ML participants. Task 2 aimed at the same information with different framing, informing participants that a machine had labeled an image with a label and asking them whether they thought the machine was correct.

Participants in Task 1 consisted of 41 raters from three different locales: US, India and Canada. Table 2 shows the number of raters from each locale. We gathered 19 ratings per image label pair (7 from American raters, 7 from Indian raters, 5 from Canadian raters), as shown in Table 3. Each rater labeled an average of 4726 image-label pairs (median 4088), with the total number of ratings provided by raters ranging from 3 to 9932. Task 1 participants were compensated monetarily in alignment with local norms of the region in which they were working.

In a subsequent validation stage, two members of the research team performed a manual review to classify the causes of model error in a sample of about 20% (2,035 image-label pairs) of the dataset from Task 1. This stage was performed in order to identify possible model error types, and categorize them for Task 2, described next.

The participants in Task 2 consisted of of 56 raters from two different locales: US and India. Table 2 shows the number of raters from each locale. We gathered 14 ratings per image label pair (7 from American raters, 7 from Indian raters). Each rater labeled an average of 2080 image-label pairs (median 1652), with the total number of ratings provided by raters ranging from 368 to 8325. Task 2 participants were compensated monetarily in alignment with local norms of the region in which they were working.

|  | Number of raters | | | |
|---|---|---|---|---|
|  | US raters | IN raters | CA raters | TOTAL raters |
| **Task 1:** Is label in image? *Annotated 10,668 image-label pairs* | 23 | 13 | 5 | 41 |
| **Validation:** Model error categorization *Annotated 2,035 image-label pairs* | 2 experts | | | |
| **Task 2:** Confirm model error *Annotated 8,326 image-label pairs* | 22 | 34 | – | 56 |

Table 1. *Size of rater pools per annotation task.*

|  | Number of raters | | | |
|---|---|---|---|---|
|  | US raters | IN raters | CA raters | TOTAL raters |
| **Task 1:** Is label in image? *Annotated 10,668 image-label pairs* | 7 | 7 | 5 | 19 |
| **Validation:** Model error categorization *Annotated 2,035 image-label pairs* | 2 experts | | | |
| **Task 2:** Confirm model error *Annotated 8,326 image-label pairs* | 7 | 7 | – | 14 |

Table 2. *Unique raters per image-label pair per annotation task.*

| Target Label | clear yes | clear no | ambiguous | TOTAL |
|---|---:|---:|---:|---:|
| Bird | 1305 | 43 | 1433 | **2781** |
| Smile | 721 | 53 | 261 | **1035** |
| Lipstick | 451 | 20 | 465 | **936** |
| Canoe | 63 | 488 | 382 | **933** |
| Chopsticks | 108 | 702 | 67 | **877** |
| Athlete | 630 | 14 | 123 | **767** |
| Muffin | 19 | 428 | 92 | **539** |
| Child | 387 | 29 | 88 | **504** |
| Chef | 32 | 214 | 138 | **384** |
| Firefighter | 69 | 70 | 160 | **299** |
| Coach | 9 | 19 | 187 | **215** |
| Construction worker | 49 | 60 | 101 | **210** |
| American football | 65 | 27 | 82 | **174** |
| Pizza | 87 | 12 | 65 | **164** |
| Selfie | 49 | 12 | 91 | **152** |
| Funeral | 24 | 23 | 98 | **145** |
| Croissant | 88 | 8 | 41 | **137** |
| Bus driver | 30 | 20 | 50 | **100** |
| Thanksgiving | 9 | 7 | 78 | **94** |
| Physician | 24 | 6 | 35 | **65** |
| Teacher | 13 | 3 | 41 | **57** |
| Graduation | 49 | 3 | 3 | **55** |
| Nurse | 19 | 3 | 23 | **45** |
| **TOTAL** | **4300** | **2264** | **4104** | **10668** |

Table 3. *Counts of how many image-label pairs for each target label fell into each supermajority vote category based on aggregated labels from raters in Tasks 1 and 2.*

**Materials**
*The CATS4ML dataset*: As described above, the CATS4ML dataset was composed of 10,668 unique submissions made by challenge participants who were instructed to identify image/label pairs within 23 pre-set labels in the OID, which they thought would be challenging for machine learning models to classify. Figure 1 shows the distribution of images across all 23 target labels - most images were submitted for the label 'bird' with an exponential long-tail distribution across all other labels.

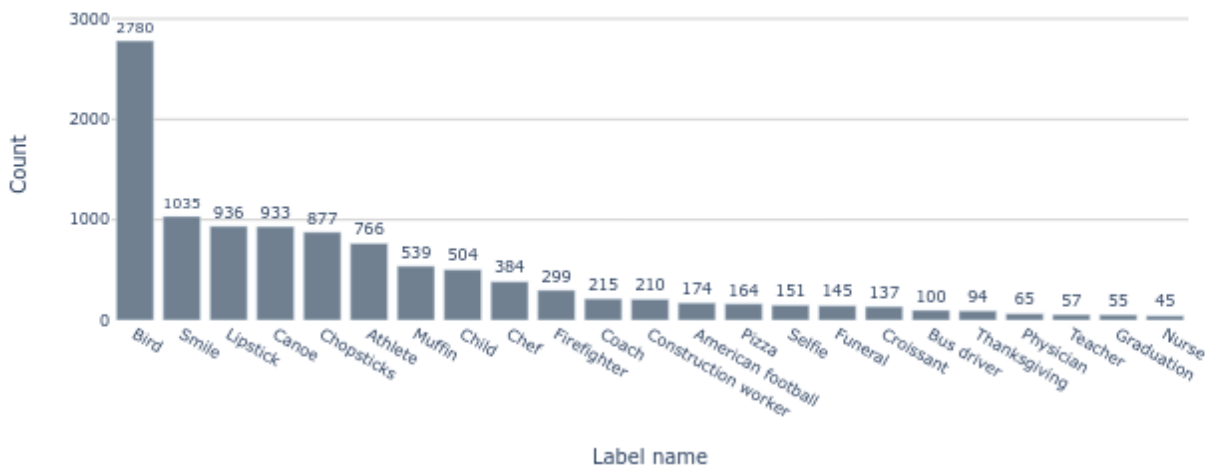

Figure 1. *Histogram of valid image/label pair counts per label name.*

*Vision models*: To provide a machine label of each image in the CATS4ML dataset, we used an ensemble of six machine vision models, each of which were state-of-the-art when they were released. These models are all Google-internal variants of the InceptionV2-based image classifier (Ioffe & Szegedy, 2015) developed in the period of 2015-2022 (including the models used in OID-V4 and OID-V3 (Krasin et al., 2017), which are publicly available through Open Images Dataset).

*Model error dataset*: Based on the findings of Task 1 and the expert qualitative validation, we constructed a subset of 8,326 image-label pairs which were then labeled in Task 2. Image/label pairs included in Task 2 met at least one of the following criteria: (i) at least one of the vision models disagreed with the human majority vote from Task 1, or (ii) there was significant disagreement among the human annotators in Task 1.

**Procedure**
*Human Annotation Task 1- Label verification*: In Task 1, participants verified whether a label applied to an image, for each of the valid image/label pairs submitted by CATS4ML participants. No specific training was provided to raters before beginning the task, as the task was injected into a general purpose image/label validation system used by the professional rater pools to perform a variety of tasks other than this one. Participants use a web interface to complete their annotation task. They worked individually, though they were able to ask questions to the experimenters along the way. In the web interface, participants view a single image and select one of three answer options indicating whether a given label applies to that image, does not apply, or they are unsure. The wording of these options is specific to the target label. If the label is "bird", as in the example in Figure 2, the options the participant chooses from are "Bird", "Not Bird", and "Unable to tell". After submitting their response to each image/label-pair, participants were immediately presented with the next image to label; they were unable to return to previous ratings to change their answer. Participants' movement through the dataset was self-paced, with an average time of 5.6 secs per image-label pair (standard deviation 1.33). In total they spent 91.2 hours on this task. All image-label pairs were annotated within 24 hours.

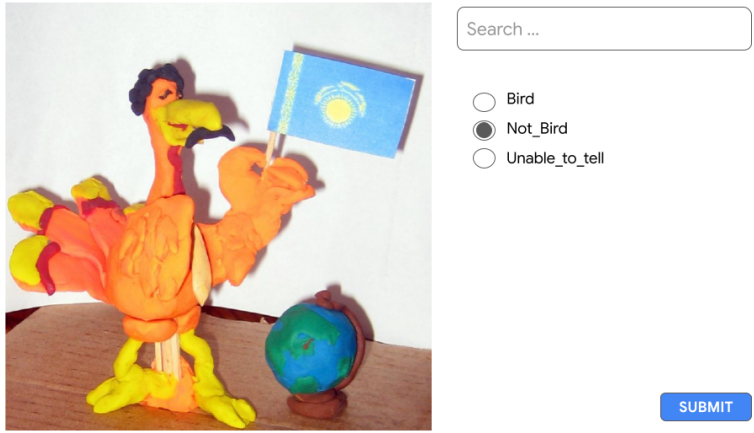

Figure 2. *Sample interface for the label verification task.*

*Human annotation Task 2- Model Error Verification*: In Task 2, participants examined specifically the *model error dataset* described above-- that is, a subset of the CATS4ML dataset that was shown in the first task and during manual validation by the researchers to in fact include image/label pairs on which human and machine labelers disagreed. Slightly differently from the first task, in Task 2 participants were informed that a model had produced a label for an image, and asked whether they thought the model was right. In cases where the human participants judged that the model had in fact made an error, they were asked to supply their reasoning as to why. The reasons that were available were as determined by expert confederates who conducted a qualitative review of about 20% of the full dataset (examples of each error reason, with images labeled as that reason, can be found in the Appendix section in Table A.1).

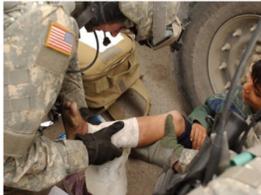

Figure 3. *Sample interface for model error verification task.*

All participants were presented guidelines on how to perform Task 2 prior to beginning. The guidelines included definitions of each of the seven categories of error. For each item, participants were presented with a link to the image, a target label for that image, and the machine's prediction for whether the target label is in the image or not (Figure 4.A). Participants answered two questions about each item: (i) whether the model indicated correctly whether the label was present in the image or not (Figure 4.B), and (ii) in the case of model error, what the type of the error was (Figure 4.C). Participants could optionally indicate a second reason for the model error. Participants could return to previous items and change their responses as needed. Participants were not given any information about how the "machine prediction" was constructed in order to avoid biasing them towards agreeing or disagreeing with the models.

**Scoring**

*Merging Task 1 and Task 2 human labels*
Both Task 1 and Task 2 in essence asked participants: "Did the machine correctly predict the label for this image?" Therefore, in order to perform data analysis collapsed across tasks, in any case in either task where the participant indicated "yes" (the machine was correct), we assert that the machine label of the image was a correct ground truth. In any case where the participant indicated "no" (the machine was not correct), we assert that ground truth is the opposite of what the machine labeled. And in any case where the participant responded "unsure", we maintain the "unsure" label as ground truth. Collapsing the data across tasks in this manner allows us to identify both image/label pairs where there was a broad consensus about whether a label applied, and also image/label pairs where even this relatively large number of raters was not able to reach a consensus. We describe the consensus schemas next; differences between tasks are also presented in the Results.

*Aggregation of human scores to supermajority vote*
We classify image-label pairs along three dimensions: (i) "clear yes" for positive examples where at least 66% of raters indicated the label was in the image, (ii) "clear no" for negative examples where at least 66% of raters indicated that the label was not in the image, and (iii) "ambiguous", for all other examples that did not meet either of the previous two criteria. Image-label pairs may fall into the ambiguous category due to either a high degree of disagreement in terms of 'yes'/'no' votes, or because of a high rate of 'unsure' answers.

**Results**

We first classify the image-label pairs that were collected in the challenge as either positive or negative examples of the submitted label. We use supermajority vote from human scores to identify which image-label pairs are positive examples ('clear yes'), negative examples ('clear no'), or could not be reliably classified due to rater disagreements or high rates of reporting that they were unsure of the correct label. Using the aggregated Task 1 and Task 2 results, we find that the challenge gathered 4300 positive examples (40.3%), 2264 negative examples (21.2%), and 4104 examples that could not be clearly classified as either positive or negative (38.5%). We present these aggregate values, broken down by the target labels, in Table 4.

| Adversariality strength: number of models fooled | Number of image-label pairs | Percent of 6554 dataset |
|---|---|---|
| 0 (not adversarial) | 1784 | 27.2 |
| 1 | 1207 | 18.4 |
| 2 | 1426 | 21.7 |
| 3 | 578 | 8.8 |
| 4 | 472 | 7.2 |
| 5 | 387 | 5.9 |
| 6 (very adversarial) | 710 | 10.8 |

Table 4. *Distribution of image-label pair adversariality across the dataset.*

Having demonstrated that the CATS4ML image/label pairs were in fact ambiguous to human raters, we next investigate whether they were ambiguous to *models*. To do this, we quantify the **adversariality of the image-label pairs** submitted to the challenge using the 61.5% of the dataset (6564 image-label pairs) on which we can compute a high-agreement human label (the 'clear yes' and 'clear no' examples in Table 4). Adversariality is computed as the number of instances of *human-model disagreements* observed across the models tested. We identify 710 (10.8%) highly adversarial image-label pairs that none of the 6 models got correct (where "correct" means "agrees with the human consensus"). This method allows us to rank the adversariality of individual images, from not adversarial through very adversarial (Table 5), based on how many models made incorrect judgements. As is evident from the table, 18.4% of images were adversarial to at least one of the state of the art models.

| Adversariality | Total pairs | Ambiguous label | Artistic depiction | Quality issue | Background context | Visual similarity | Out of context | Atypical depiction | Other error reason |
|---|---|---|---|---|---|---|---|---|---|
| 1 | 1137 | 7 | 6 | 1 | 10 | 45 | 0 | 15 | 13 |
| 2 | 1404 | 36 | 21 | 22 | 570 | 207 | 6 | 832 | 549 |
| 3 | 562 | 16 | 7 | 5 | 216 | 102 | 0 | 308 | 156 |
| 4 | 471 | 18 | 5 | 6 | 172 | 144 | 4 | 228 | 107 |
| 5 | 387 | 14 | 5 | 1 | 114 | 196 | 2 | 151 | 85 |
| 6 | 710 | 26 | 15 | 10 | 212 | 389 | 0 | 234 | 125 |

Table 5. *Total image-label pairs for which a given error reason was indicated by at least 25% of raters in the Task 2 qualitative labeling task. Totals are different from Table 4 because only a subset of the full CATS4ML dataset was rated in Task 2. "Total pairs" represents the total number of image-label pairs that were rated in Task 2. Totals across rows may be greater than the total number of image-label pairs as examples can be labeled with more than one error reason.*

Next, we investigate the *reasons* for the high disagreement between humans, observed in the 38% of image-label pairs where there was no supermajority agreement. For this, we consider the full CATS4ML dataset of 10668 items, and we assess ambiguity from three perspectives: *disagreements due to rater characteristics*, *disagreements due to the image-label pairs*, and *disagreements due to the rating task set up*. These three perspectives have previously been identified as relevant to understanding crowd labels and rater disagreements (Aroyo & Welty, 2014).

We begin by constructing a *mixed-effects model* to examine how these factors explain variance in the data. We construct a null model predicting whether the rater indicated that the label is in the image or not (i.e., 'yes' or 'not yes'), with random intercepts for raters and items. We compare this null model to three separate models that add fixed effects of (i) rater locale, (ii) label id, and (iii) task type. We compare each of the three predictor models to the null model using an ANOVA. Table 6 shows that each of these three models is a significantly better fit for the data compared to the null model, indicating that rater characteristics (as indexed by locale), the label name, and the task framing all explain a significant degree of variance in the data. To determine whether these three factors interact with each other, we construct both an *additive and an interactive model* using all three predictors and observe that the interactive model is a significantly better fit for the data compared to the additive model ($p < 0.001$).

| Model description | Model definition | AIC | BIC | Fit compared to null model |
|---|---|---|---|---|
| Null model baseline | Rating ~ 1 + (1|rater_id) + (1|item_id) | 289711.9 | 289754.4 | N/A |
| Rater locale model | Rating ~ Locale + (1|rater_id) + (1|item_id) | 289677.0 | 289740.7 | $p < 0.001$ |
| Task type model | Rating ~ Task_type + (1|rater_id) + (1|item_id) | 289669.7 | 289722.8 | $p < 0.001$ |
| Label name model | Rating ~ Label_name + (1|rater_id) + (1|item_id) | 282069.8 | 282324.5 | $p < 0.001$ |
| Additive model (all predictors) | Rating ~ Locale + Label_name + Task_type + (1|rater_id) + (1|item_id) | 282007.2 | 282293.8 | $p < 0.001$ |
| Interactive model (all predictors) | Rating ~ Locale * Label_name * Task_type + (1|rater_id) + (1|item_id) | 271579.6 | 272725.9 | $p < 0.001$ |

Table 6. *Mixed effect model definitions and fit statistics.*

These three factors interact in complex ways, and interpretation of which levels of each factor affect the rating behavior of participants is less central to the investigation of sources of disagreement than systematically investigating each of these factors individually. The interpretation of these interactions would each be dependent on the reference level chosen, and in the case of the Label name predictor, there is no sensible reference value to choose, as each label is completely independent of the others. We thus use these results to justify further investigation of each of the three sources of disagreement, but note the lack of analysis of interactions between these factors as a limitation.

*Disagreements due to rater characteristics*
For both Task 1 and Task 2, we investigate agreement between raters with Krippendorf's alpha (inter-rater reliability; IRR) and cross-replication reliability (xRR).  Overall agreement was only moderate in both tasks (Task 1:  0.46, Task 2: 0.20).  In Task 1, IRR was higher within locale than across locale, and xRR revealed that Indian and American raters agreed with each other more than did Indian and Canadian raters or Canadian and American raters.  Task 1 results are summarized in Table 7. In Task 2, agreement was even poorer.  Task 2 results are summarized in Table 8.  Taken together, these results make clear that human labelers did not, as a rule, agree with each other on how the images in the CATS4ML dataset should be labeled, and, that a rater's locale did have an impact on how that rater labeled images.  Table 9 provides examples of images where different locales reached different consensus labels.  Panel (b) presents an image where 92% of American raters affirmed that the label "bird" was appropriate while 86% of Indian raters were unsure.

| Metric | Locale | Agreement |
|---|---|---|
| IRR | OVERALL | 0.4737 |
| | IN | 0.5739 |
| | US | 0.5349 |
| | CA | 0.3794 |
| xRR | IN x US | 0.5429 |
| | IN x CA | 0.4653 |
| | US x CA | 0.5088 |

Table 7. *IRR and xRR scores for Task 1, by rater locale.*

| Metric | Locale | Agreement |
|---|---|---|
| IRR | OVERALL | 0.1982 |
| | IN | 0.3624 |
| | US | 0.1299 |
| xRR | IN x US | 0.1846 |

Table 8. *IRR and xRR scores for Task 2, by rater locale.*

| | a) 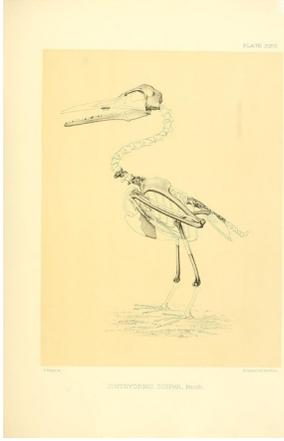  **Label:** BIRD  **Human majority:** 'unsure' | | | b) 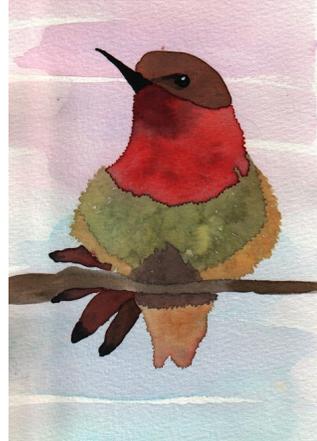  **Label:** BIRD  **Human majority:** 'yes' | | |
|---|---|---|---|---|---|---|
| | Yes % | Unsure % | No % | Yes % | Unsure % | No % |
| US raters | **67** | 17 | 17 | **92** | 0 | 8 |
| CA raters | 20 | 20 | **60** | 40 | **60** | 0 |
| IN raters | 0 | **100** | 0 | 0 | **86** | 14 |

Table 9. *Examples of images where the raters in different locales respond differently when asked if the label is in the image.*

*Disagreements due to the image-label pairs*

To this point, we have examined image/label pairs where the human raters were able to make a judgment about whether the label was in the image. Next, we examine image/label pairs where humans responded that they were "unsure" if the label was in the image. We identified 2039 examples (21.5% of all image-label pairs) in which the 'unsure' label was the most frequently selected label across raters in Task 1. We provide two illustrative examples in Table 10 from *events* and *professions* label types. In the first case, where the label given is THANKSGIVING, it is genuinely ambiguous whether the meal is a Thanksgiving dinner, and in the second it is also ambiguous whether the people wearing white coats are PHYSICIANs, as opposed to any other profession that wears a lab coat. In both cases, the label is *potentially* consistent with the image, but crucially disambiguating background information about the image's setting, date, or participants is unavailable to the raters.

| | a) 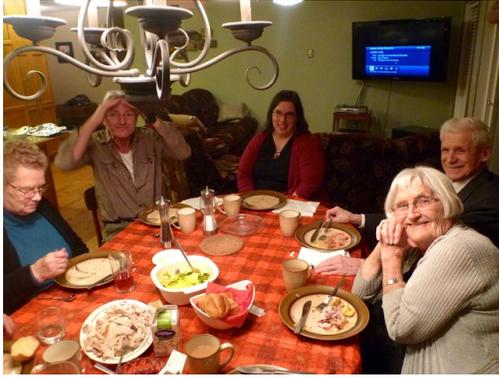 **Label:** THANKSGIVING; **Human majority:** 'unsure' | | | b) 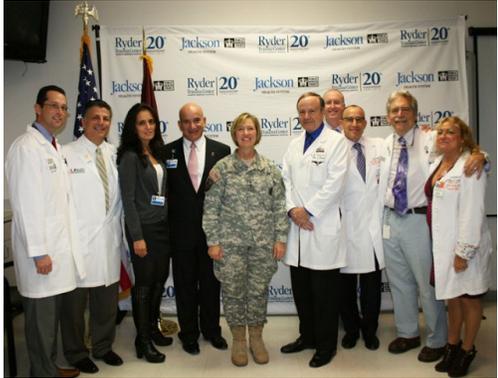 **Label:** PHYSICIAN; **Human majority:** 'unsure' | | |
|---|---|---|---|---|---|---|
| | Yes % | Unsure % | No % | Yes % | Unsure % | No % |
| US raters | 42 | **50** | 8 | 25 | **50** | 25 |
| CA raters | 40 | **60** | 0 | 20 | **60** | 20 |
| IN raters | 25 | **75** | 0 | 29 | **57** | 14 |

Table 10. *Examples of images where the majority of humans indicate they are UNSURE if the label is in the image.*

The labels in this study spread across a range of different types of concepts: concrete, abstract, events, roles, and professions. Some of these categories are inherently more difficult in an image-labeling task. Both professions and roles can be heavily context-dependent. Events can be difficult to determine from a single image, as many types of events include multiple sub-parts to the whole (e.g., is 'Thanksgiving' just a nicely-dressed turkey?).

*Disagreements due to the rating task*

To compare results of Task 2 with those of Task 1, we first compute the rate of positive, negative, and ambiguous image-label pairs for each label, subsetting the Task 1 results to just the image-label pairs that appeared in both tasks. We observe some differences between the two tasks, with 35.8% of the image-label pairs switching supermajority vote labels between Task 1 and Task 2. The cross-task comparison is presented in Table 11.

| Supermajority vote label | | Number of examples | Percent of total |
|---|---|---|---|
| Task 1: Is label in image? | Task 2: Is machine correct? | | |
| Yes | **Yes** | **2714** | **32.6** |
| | *No* | *6* | *0.1* |
| | *Ambiguous* | *464* | *5.8* |
| No | *Yes* | *9* | *0.1* |
| | **No** | **845** | **10.2** |
| | *Ambiguous* | *614* | *7.4* |
| Ambiguous | *Yes* | *1561* | *18.8* |
| | *No* | *325* | *3.9* |
| | **Ambiguous** | **1787** | **21.5** |

Table 11. *Cross Task comparison. In **bold** are rows representing image/label pairs that had consistent supermajority labels across tasks. In italics are rows representing image/label pairs that had super inconsistent supermajority labels across tasks.*

For the over one third of image-label pairs that *flip their label based on the task phrasing,* most of these flips involve the 'ambiguous' label that we apply when aggregating to supermajority vote, indicating that there are relatively few cases where raters truly change their vote from 'yes' to 'no' (or vice versa). To illustrate these cases, we randomly select an image-label pair from each of the six different kinds of label flips observed, and show the examples along with the raters' labeling patterns in Tables 12, 13 and 14. It is difficult to draw very general conclusions about these images, as we observe patterns where the human supermajority vote label switches to align with the machine label shown in Task 2 (12a, 13a, 14b) and to contradict the machine label shown (14a). However, these images are illustrative of the kinds of difficulties that participants had in assigning labels, and they show that slight changes in the wording or presentation of the task can lead to different results, even on a task that appears straightforward.

|  | **Task 1 label:** Yes | |
|---|---|---|
|  | **Task 2 label:** No | **Task 2 label:** Ambiguous |
|  | a) 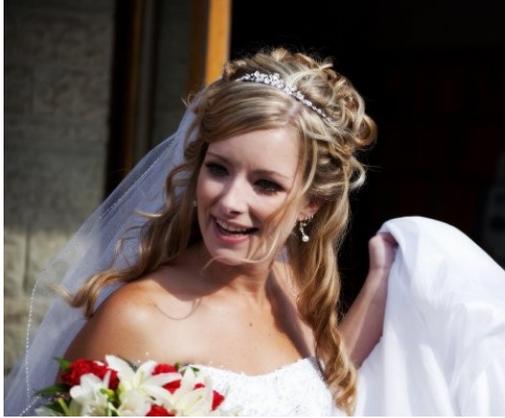 **Label:** LIPSTICK **Task 2 Machine label:** No | b) 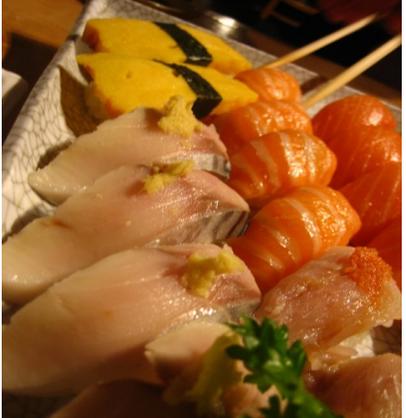 **Label:** CHOPSTICKS **Task 2 Machine label:** No |
|  | Yes % | Unsure % | No % | Yes % | Unsure % | No % |
| Label-in-image | **68.4** | 0.0 | 31.6 | **84.2** | 10.5 | 5.3 |
| Confirm-model-error | 28.6 | 0.0 | **71.4** | **64.3** | 0.0 | 35.7 |

Table 12. *Examples of images where the supermajority vote label was different between the two tasks, focusing on examples that flipped an original 'yes' label in the Label-in-Image task.*

|  | **Task 1 label:** No | |
|---|---|---|
|  | **Task 2 label:** Yes | **Task 2 label:** Ambiguous |
|  | a) 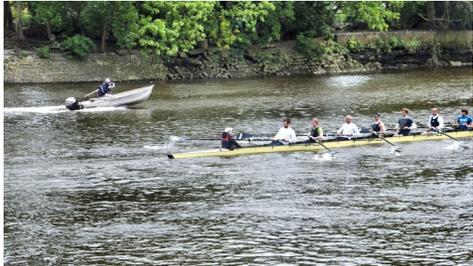 **Label:** CANOE **Task 2 Machine label:** Yes | b) 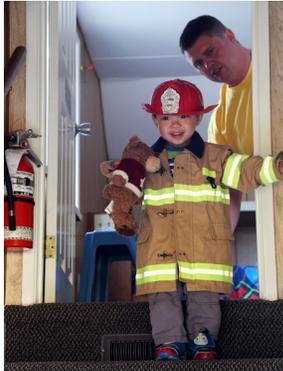 **Label:** FIREFIGHTER **Task 2 Machine label:** Yes |
|  | Yes % | Unsure % | No % | Yes % | Unsure % | No % |
| Label-in-image | 15.8 | 10.5 | **73.7** | 15.8 | 15.8 | **68.4** |
| Confirm-model-error | **71.4** | 7.1 | 21.4 | **50.0** | 14.2 | 35.7 |

Table 13. *Examples of images where the supermajority vote label was different between the two tasks, focusing on examples that flipped an original 'no' label in the label-in-image task.*

|  | **Task 1 label:** Ambiguous ||
|  | **Task 2 label:** Yes | **Task 2 label:** No |
|  | a) 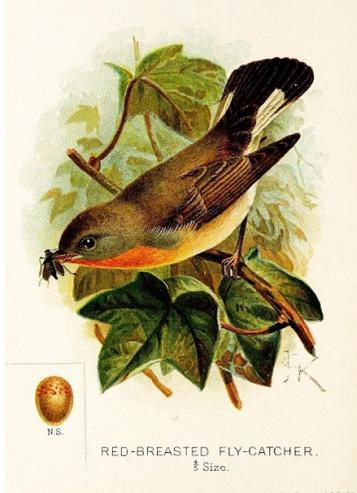<br>**Label:** BIRD<br>**Task 2 Machine label:** No | b) 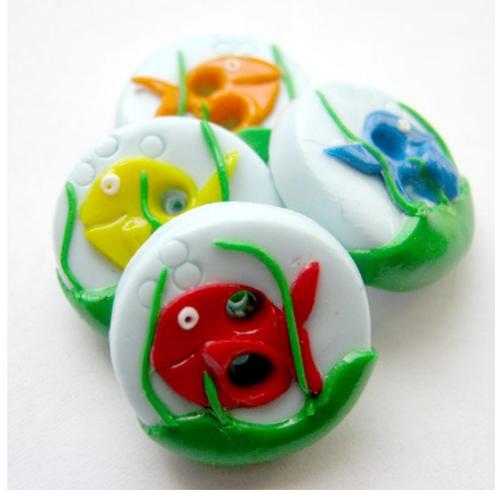<br>**Label:** SMILE<br>**Task 2 Machine label:** No |
|  | Yes % / Unsure % / No % | Yes % / Unsure % / No % |
| Label-in-image | 31.6 / **63.2** / 5.3 | 0.0 / **52.6** / 47.6 |
| Confirm-model-error | **92.9** / 0.0 / 7.1 | 7.1 / 0.0 / **92.9** |

Table 14. *Examples of images where the supermajority vote label was different between the two tasks, focusing on examples that flipped an original 'ambiguous' label in the label-in-image task.*

**Discussion**

Here, we were concerned with the problem of label ambiguity in large label space models. We aimed to better understand whether and how humans and machines disagree with each other about how training data for large label space models should be assigned a ground truth. As we introduced, label ambiguity is typically deleterious to model performance, meaning that, when a model is to be created to serve a public good, label ambiguity should be avoided to the extent possible.

We demonstrated that it is, in fact, challenging for human raters and machines to agree with each other on label ground truth, even for relatively concrete concepts such as "bird" and "lipstick." We further demonstrated that the geographical location in which a human rater is situated can have an impact on how that rater performs the labeling task. And, we demonstrated that small changes to the way a labeling task is framed can also have an impact (that interacts with locale) on how the task will be performed. Given all these potential complications to performing the bedrock task of machine vision model training (assigning ground truth to images), we conclude with our recommendations as to how developers and policy makers can best address label ambiguity.

First, take a community-driven approach to data labeling. Make sure that the people doing the labeling are from the communities that are going to be impacted by the model deployment. Second, assume that there will always be variance, ambiguity, and subjectivity in any data labeling task, regardless of how simple it may seem on its face. Embrace the fact that there is not, and cannot be, one singular 'gold standard'. To the extent that it is possible, attempt to identify and explore possible sources of ambiguity in any data set, and understand how these sources of ambiguity might be related to the communities impacted by the model. Finally, define and deploy metrics that can measure ambiguity in data. For example, if data are labeled in different sessions, on different interfaces, or by different pools raters, measure and track differences between data subsets. Slice data by demographic properties of the community that will be impacted by the data, such as geographic location (as was done here), gender, age, or ability. Measure and track any differences across slices.

Adopting these recommendations will ensure that, at a minimum, a deployed model has been contributed to by the community it serves, that possible sources of model failure are understood and tracked, and that the manner in which the model is serving different subsets of the community is also understood and tracked. A model deployed under these conditions is on the right track to responsibly serve its community.

**Acknowledgements**
We thank Praveen Paritosh, Ka Wong and Samia Ibtasam for their work in co-organizing the CATS4ML challenge and initial data analysis with Lora Aroyo. We thank Christina Greer for providing a pre-publication review of this paper.

**Appendix**

| Error reason | Super-majority vote | Task 2 Machine label | Percent of raters | Image |
|---|---|---|---|---|
| Artistic depiction of the label | Task 1: Ambiguous<br>Task 2: Yes | No | 78.6 | **Label: BIRD**<br>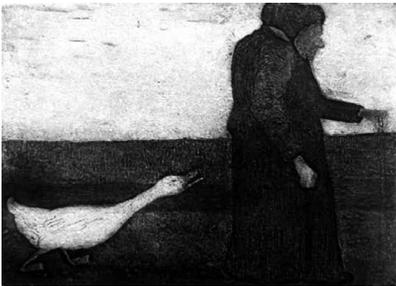 |
| Machine over-relied on background context | Task 1: Ambiguous<br>Task 2: Yes | No | 85.7 | **Label: BIRD**<br>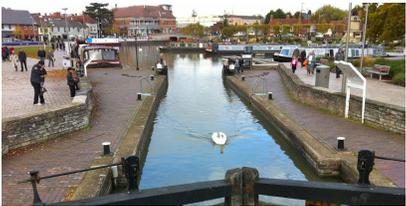 |
| Object is depicted out of typical context (e.g., no background) | Task 1: Yes<br>Task 2: Yes | No | 35.7 | **Label: ATHLETE**<br>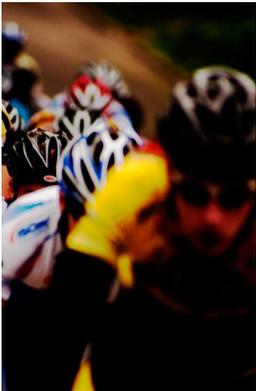 |
| Unexpected or atypical depiction of the label | Task 1: Ambiguous<br>Task 2: Ambiguous | No | 71.4 | **Label: CHILD**<br>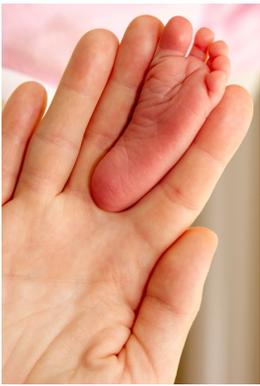 |
| Ambiguous meaning of the label (e.g. triggers different interpretation) | Task 1: Ambiguous<br><br>Task 2: No | Yes | 35.7 | **Label: CONSTRUCTION WORKER**<br>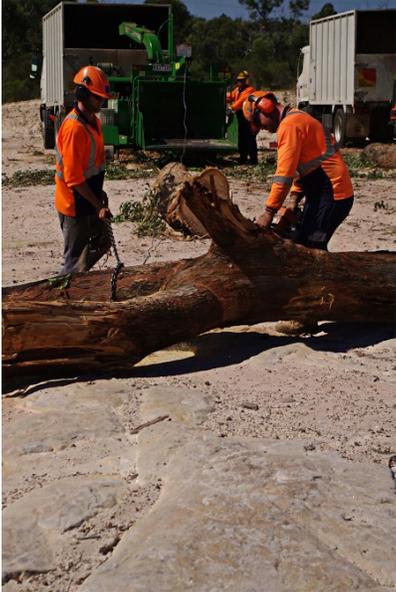 |
| Visually similar shape of the label | Task 1: No<br>Task 2: No | Yes | 85.7 | **Label: MUFFIN**<br>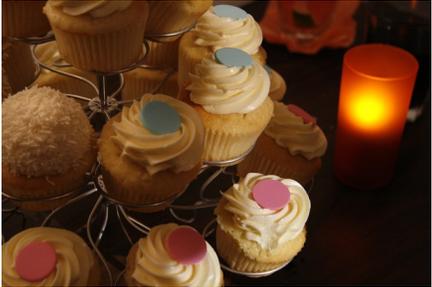 |
| Image has quality issue | Task 1: No<br>Task 2: No | Yes | 64.3 | **Label: SELFIE**<br>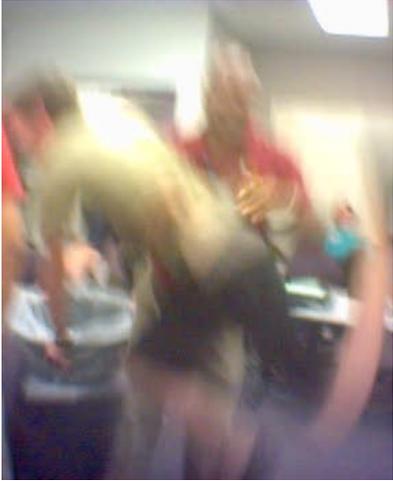 |

| OTHER reason for model error | Task 1: Yes<br>Task 2: Yes | No | 64.3 | 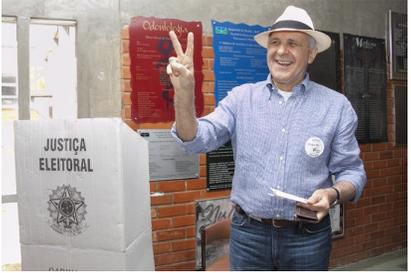 |

Table A.1. *All error reasons from Task 2. Percent of raters indicates the percentage of Task 2 participants who indicated that the model was wrong for that particular error reason, either as the primary or secondary reason for the model error.*